\documentclass[12pt]{iopart}

\usepackage{xcolor}
\usepackage{latexsym}

\begin{document}

\title{Graviton-photon oscillation in alternative theories of gravity}

\author{Jos\'e A.~R.~Cembranos, Mario Coma D\'iaz, and\\ Prado Mart\'in-Moruno}

\address{Departamento de F\'isica Te\'orica and UPARCOS, \\Universidad Complutense de Madrid,
E-28040 Madrid, Spain}
\ead{cembra@ucm.es, marioctk@gmail.com, pradomm@ucm.es}
\vspace{10pt}

\begin{abstract}
In this paper we investigate graviton-photon oscillation in the presence of an external magnetic field in alternative theories of gravity. 
Whereas the effect of an effective refractive index for the electromagnetic radiation was already considered in the literature, we develop the first approach to take into account the effect of the modification of the predictions for gravitational waves in alternative theories of gravity in the phenomenon of graviton-photon mixing.
\end{abstract}




\section{Introduction}

We are witnessing the dawn of gravitational wave (GW) astronomy. Most researchers in our field are probably wondering about how this new window to our Universe will affect our theoretical knowledge. It is hight time to dust off seminal studies on GWs and get them in shape taking into account recent developments in gravitation. An optimal example is the phenomenon of graviton-photon oscillation, which has long been known by the general relativistic community. 
In 1960 Gertsenshtein investigated the excitation of GWs during the propagation of electromagnetic waves (EMWs) in an electric or magnetic field~\cite{Gertsenshtein}. Later on, Lupanov considered the inverse process, that is EMWs creation by GWs crossing an electric field, as an indirect way of measuring GWs~\cite{Lupanov}. 
Moreover, he argued that the effectiveness of a hypothetical detector could be improved by introducing a dielectric material with high permittivity~\cite{Lupanov}. As it was later investigated in more detail by Zel'dovich, the change in the speed of light due to an effective refractive index induces loss of coherence and, therefore, affects the conversion process. The formalism of photon mixing with gravitons (and with low-mass particles  in general) was completely developed by Raffelt and Stodolsky~\cite{Raffelt:1987im} taking into account the effective refractive index for EMWs of quantum-electrodynamic origin. 
It should be emphasized that graviton-photon mixing could provide us with an indirect way to measure GWs. Although direct detection of GWs is already a reality, our ability to detect EMWs is still much better by far. Whereas the effect produced by conversion of EMWs in GWs and back again is undetectable by laboratory experiments~\cite{Boccaletti}, it has been pointed out that graviton-photon mixing of primordial GWs in the presence of cosmic magnetic fields could lead to observable footprints in the X-ray cosmic background~\cite{Dolgov:2012be}.

On the other hand, nowadays alternative theories of gravity (ATGs) are being analysed from different perspectives, since they can easily describe an accelerating universe. Indeed, Starobinsky's inflationary model is considered to be one of the most promising scenarios to describe the early universe \cite{Ade:2015xua}.
As general relativity (GR), ATGs predict the emission and propagation of gravitational radiation. The nature of this gravitational radiation, however, can be very different and could have more than two independent polarizations~\cite{Will:2014kxa}. Focusing attention on the two tensor modes, that is the GW, their production and propagation is typically modified with respect to the general relativistic predictions~\cite{deRham:2014zqa,Saltas:2014dha,Bettoni:2016mij,Caprini:2018mtu,Akrami:2018yjz}.
In fact, as it has been recently shown explicitly, one could consider that GWs propagate in the framework of ATGs as if they were in a diagravitational medium characterized by three different constitutive tensors, reducing just to an effective refractive index when considering perturbations around highly symmetric background spacetimes~\cite{Cembranos:2018lcs}. 
Hence the possibility of using GWs observations to constrain ATGs has been taken into account seriously, as they could provide us with an \emph{experimentum crucis} to discern what is the theory of gravity describing our Universe. Indeed, the recent measurement of GWs and their electromagnetic counterpart~\cite{TheLIGOScientific:2017qsa} has allowed to discard  the relevance during the late universe of modifications of GR generated by a large number of ATGs~\cite{Lombriser:2015sxa,Lombriser:2016yzn,Ezquiaga:2017ekz,Creminelli:2017sry,Sakstein:2017xjx,Baker:2017hug,Akrami:2018yjz}, singling out some theories promising as dark energy mimickers~\cite{Baker:2017hug,Belgacem:2017cqo,BeltranJimenez:2018ymu}. Anyway, those modifications could still play a relevant role during the early universe.

Following this spirit, in this paper we focus our attention on the phenomenon of graviton-photon mixing, which is also known as graviton-photon oscillation, and investigate how it is modified by a diagravitational refractive index  and an effective gravitational coupling constant. In our treatment, we shall firstly take a particular family of ATGs for concreteness, which is Horndeski's theory~\cite{Horndeski:1974wa}; however, we will generalize our results for more general ATGs. As we shall show, the graviton-photon mixing probability depends on the effective gravitational coupling of the tensors modes and the effective (electromagnetic and gravitational) refractive indexes.
Up to the best of our knowledge, this is the first attempt to address graviton-photon oscillation in the framework of ATGs. Nevertheless, we would like to mention some interesting works in related topics. GWs-gauge field oscillations have been recently investigated~\cite{Caldwell:2016sut,Caldwell:2017sto} and its potential observational consequences anbalysed~\cite{Caldwell:2018feo}. In addition, there has been developments in understanding how the background can affect the propagation of GWs in GR~\cite{Flauger:2017ged} and also in $f(R)$-gravity~\cite{Bamba:2018cup}. On the other hand, the phenomenon of photon production in a GW background seeded by vacuum fluctuations instead of a background classical electromagnetic field has been scrutinized in references~\cite{Jones:2015uda,Jones:2017dzt} and references therein. Such phenomenon should also be predicted in ATGs that can describe the same background in the linearized limit as GR.

This paper can be outlined as follows: In section~\ref{sec:mixing} we review the phenomenon of graviton-photon mixing in a general relativistic background. In section~\ref{sec:MG} we investigate this mixing in the framework of ATGs, although we give special emphasis to Horndeski theory our results go much beyond that case. In section~\ref{sec:P} we show how the probability of conversion of GWs into EMWs is modified for ATGs. Finally, in section~\ref{sec:d}, we discuss our results.

\section{Graviton-photon mixing}\label{sec:mixing}
Let us start by briefly summarizing the description of graviton-photon mixing in GR. Raffelt and Stodolsky argued that the quantum corrections to the electrodynamic system could be relevant for graviton-photon oscillation in regions of strong magnetic fields~\cite{Raffelt:1987im} (see also reference~\cite{Dolgov:2012be}). Those corrections, which can be taken into account adding an Euler--Heisenberg term into the Lagrangian, lead to an effective refractive index that the photon experience even in vacuum. Apart from this term, the the electromagnetic Lagrangian takes the usual form
\begin{eqnarray}
\mathcal{L}_{\rm EM}=-\frac{1}{4}F_{\mu\nu}F^{\mu\nu}
\end{eqnarray}
where $F^{\mu\nu}$ is the electromagnetic field tensor.
On the other hand, it is useful to consider that the metric perturbations, defined through
\begin{equation}\label{gh}
g_{\mu\nu}=\eta_{\mu\nu}+\kappa^2h_{\mu\nu},
\end{equation}
are expressed in the transverse-traceless gauge. Therefore, the quadratic gravitational Lagrangian in tensor perturbations and the interaction term with the electromagnetic field can be written as
\begin{equation}
\mathcal{L}_{\rm GR}= -\frac{1}{4}\partial_\mu h_{ij} \partial^\mu h^{ij}, \quad {\rm and} \quad \mathcal{L}_{\rm int}=\frac{\kappa}{2}\,h_{ij}T^{ij},
\end{equation}
with $T^\mu{}_\nu=F^{\mu\rho}F_{\nu\rho}-\delta^\mu_\nu F_{\alpha\beta}F^{\alpha\beta}/4$. Note that the total electromagnetic field is the sum of the external magnetic field, that is $B^i_{\rm e}$, which is considered to be static and homogeneous, and the EMW, $B^j$, which field strenght is much weaker. Therefore, focusing attention on the wave-form contributions, we have $h_{ij}T^{ij}=-h_{ij}B^i_{\rm e}B^j$ and the graviton-photon dynamical system takes the form
\begin{eqnarray}
\qquad\quad\Box h_{ij}&=&\kappa\,B_i^{\rm (e)}B_j,\label{pmetric}\\
\left(\Box-m^2_{\gamma}\right)A^j&=&\kappa\,\partial_ih^{jk}F_k^{{\rm (e)}i}\label{pvector},
\end{eqnarray}
where $A^j$ is the vector potential of the electromagnetic wave and $m_{\gamma}$ is the effective photon mass implied by the effective refractive index~\cite{Dolgov:2012be}.
We fix the $z$ axis along the direction of propagation of the waves without loss of generality. Then, the gravitational wave tensor and the electromagnetic vector potential can be expressed as
\begin{eqnarray}
h_{ij}(t,\,z)&=&\left[h_+(z)\,\epsilon^+_{ij}+h_\times(z)\,\epsilon^\times_{ij}\right]e^{-i\omega t},\label{hp}\\
A_j(t,\,z)&=&i\left[A_x(z)\,\epsilon^x_j+A_y(z)\,\epsilon^y_j\right]e^{-i\omega t},
\end{eqnarray}
where the polarization tensors have non-vanishing componentes $\epsilon^+_{xx}=-\epsilon^+_{yy}=1$ and $\epsilon^\times_{xy}=\epsilon^\times_{yx}=1$, and we have introduced a phase in the vector for convenience, as it was explicitly done in reference~\cite{Dolgov:2012be}. Taking into account the form of the expansion~(\ref{hp}) in the interaction term $h_{ij}B^i_{\rm e}B^j$, one can conclude that only the external magnetic field transverse to the direction of the wave propagation contribute to the phenomenon, that is $B_{\rm T}=B_{\rm e}\sin\phi$ with $\cos\phi=\hat B_{\rm e}\cdot\hat k$. Assuming for simplicity that the transverse magnetic field lies along the $y$-direction, taking $k= i\partial_z$, so that $k_{\gamma}\,A=i\partial_zA$ and $k_{\rm g}h=i\partial_zh$, and noting that $|n_{\gamma}-1|\ll1$ with $k_{\gamma}=n\,\omega$, it is easy to recover the result of Raffelt and Stodolsky. This is~\cite{Raffelt:1987im}
\begin{eqnarray}
\left[\left(\omega-i\partial_z\right)+
\left({\begin{array}{cccc}
   \Delta^\bot_{\gamma} & \Delta_M &0&0 \\
   \Delta_M & 0 &0&0\\
     0&0&\Delta^\parallel_{\gamma} & \Delta_M \\
   0&0&\Delta_M & 0 \\
  \end{array} } 
\right)\right]
\left(\begin{array}{c}
A_\bot\\h_+\\A_\parallel\\h_\times
\end{array}\right)=0,
\end{eqnarray}
with $A_\bot$ and $A_\parallel$ denoting the components of the wave vector field perpendicular and parallel to the transverse magnetic field. The mixing term depends on the gravitational coupling and on the transverse magnetic field $\Delta_M=\kappa\, B_{\rm T}/2$. On the other hand, the effective photon mass term has led to a modification in the propagation of EMWs that can be different for both polarizations (as it has been argued to be the case~\cite{Raffelt:1987im}); that is $\Delta^{\bot, \parallel}_{\gamma}=\omega\, (n_{\gamma}^{\bot, \parallel}-1)$.

\section{Mixing in a diagravitational medium}\label{sec:MG}
Now, let us firstly focus in a particular  family of theories and show later how our results apply to several ATGs. So, for concreteness we start studying Horndeski theory, which deals with the most general scalar-tensor Lagrangian leading to second order equations of motion~\cite{Horndeski:1974wa,Deffayet:2011gz}. 
This is~\cite{Kobayashi:2011nu}
\begin{eqnarray}
\hspace{-2cm}\mathcal{L}_{\rm H}&=&
K(\phi, X)
-G_3(\phi, X)\Box\phi, 
+G_{4}(\phi, X)R+G_{4X}\left[\left(\Box\phi\right)^2-\left(\nabla_\mu\nabla_\nu\phi\right)^2\right]\nonumber\\
\hspace{-2cm}&+&G_5(\phi, X) G_{\mu\nu}\nabla^\mu\nabla^\nu\phi-\frac{G_{5X}}{6}\Bigl[\left(\Box\phi\right)^3-  3\left(\Box\phi\right)\left(\nabla_\mu\nabla_\nu\phi\right)^2+ 2\left(\nabla_\mu\nabla_\nu\phi\right)^3\Bigr],
\end{eqnarray}
with $X= - \partial_{\mu} \phi \partial^{\mu} \phi / 2$ being the scalar-field kinetic term, $G_{i X}=\partial G_{i} / \partial X $, $(\nabla_\mu\nabla_\nu\phi)^2=\nabla_\mu\nabla_\nu\phi\nabla^\mu\nabla^\nu\phi$, and
$(\nabla_\mu\nabla_\nu\phi)^3=\nabla_\mu\nabla_\nu\phi\nabla^\nu\nabla^\lambda\phi\nabla_\lambda\nabla^\mu\phi$. 
We have in mind the potential relevance of graviton-photon mixing in cosmological scenarios, although when dealing with sub-Hubble modes the background geometry can be approximated by Minkowski space. Therefore, we assume that the scalar field is homogenous, that is $\phi=\phi(t)$. Then, the quadratic Horndeski Lagrangian for the metric perturbation~\cite{Kobayashi:2011nu} can be written as
\begin{equation}\label{Hp}
\mathcal{L}_{\rm H}=\frac{\kappa^2}{4\kappa_{\rm eff}^2(\phi,\,X)}\left[\dot h_{ij}^2-c_T^2(\phi,\,X)(\nabla h_{ij})^2\right],
\end{equation}
where
\begin{eqnarray}
\kappa_{\rm eff}(\phi,\,X)&=&1/\sqrt{G(\phi,\,X)},\quad {\rm and}\quad c_T^2=\frac{F(\phi,\,X)}{G(\phi,\,X)},
\end{eqnarray}
with
\begin{eqnarray}
F(\phi,\,X)&=&G_4-X\left(\ddot\phi\, G_{5X}+G_{5\phi}\right),\\
G(\phi,\,X)&=&G_4-2XG_{4X}+XG_{5\phi},
\end{eqnarray}
and one recover GR for $\kappa_{\rm eff}=\kappa$.
As we want to focus our attention on the mixing of gravitons with photons, we neglect the backreaction of the scalar field into the background geometry when using expansion (\ref{gh}). In addition, we will also neglect the effect of the coupling of the scalar field perturbations to the GWs, which can appear in generic situations, see for example \cite{Bettoni:2016mij}. Under the mentioned approximations and assuming again that the electromagnetic field is minimally coupled to gravity, the first equation for the graviton-photon system in the presence of an external magnetic field, that is equation (\ref{pmetric}), is modified. Now it reads
\begin{eqnarray}
\left\{\Box h_{ij}-\left[\nu\,\partial_0-\left(1-c_T^2\right)\partial_z^2\right]\right\}=\frac{\kappa_{\rm eff}^2}{\kappa}\,B_i^{\rm (e)}B_j,\label{pmetricH}
\end{eqnarray}
where $\nu=-2\,\dot \kappa_{\rm eff}/\kappa_{\rm eff}$. We apply similar arguments to those used in the general relativistic case~\cite{Raffelt:1987im,Dolgov:2012be} reviewed in the previous section.  It simplifies the treatment to consider that the modification on the propagation of GWs can be encapsulated in an effective diagravitational refractive index \cite{Cembranos:2018lcs} 
\begin{eqnarray}
n_{\rm g}=\frac{1}{c_T}\sqrt{1+i\frac{\nu}{\omega}}.
\end{eqnarray}
For simplicity, we will consider that $\nu$ and $c_T$ are approximately constant during the interval of interest.
Note that, even if the diagravitational refractive index could significantly separate from unity at early cosmic epochs, we will assume that both $|n_{\gamma}-1|\ll1$ and $|n_{\rm g}-1|\ll1$. Therefore, we have \cite{Cembranos:2018lcs}
\begin{eqnarray}\label{neff}
n_{\rm g}=1+i\frac{\nu }{2\,\omega}+(1-c_{T}),
\end{eqnarray}
up to first order.

Now, let us note that ATGs commonly produce a refractive index for GWs propagating in a Minkowski background and/or have an effective gravitational coupling $\kappa_{\rm eff}$. On one hand, the perturbed Lagrangian (\ref{Hp}) is also compatible with other ATGs different from Horndeski's theory. On the other hand, ATGs based on a massive graviton or on Lorentz violations can produce other terms in the index (\ref{neff}), leading to~\cite{Cembranos:2018lcs}
\begin{equation}\label{neff2}
n_{\rm g}=1+i\frac{\nu }{2\,\omega}+(1-c_{T})-\frac{m_{\rm g}^2}{2\,\omega^2}-\frac{A}{2}\omega^{\alpha-2},
\end{equation}
up to first order in $\nu/\omega$, $1-c_T$, $m_g^2/\omega^2$, and $A\,\omega^{\alpha-2}$. In addition, some theories can produce birrefrigence effects, that is $n_{\rm g}^\bot\neq n_{\rm g}^\parallel$ \cite{Alexander:2004wk,Yunes:2010yf}. So, we take into account this more general form of the refractive index and follow the procedure outlined in the previous section to obtain the mixing matrix for more general ATGs. The graviton-photon system is, therefore, given by
\begin{eqnarray}\label{Matrix}
\left[\left(\omega-i\partial_z\right)+
\left({\begin{array}{cccc}
   \Delta^\bot_{\gamma} & \Delta_M &0&0 \\
   \Delta_M & \Delta_{\rm g}^\bot &0&0\\
     0&0&\Delta^\parallel_{\gamma} & \Delta_M \\
   0&0&\Delta_M & \Delta_{\rm g}^\parallel \\
  \end{array} } 
\right)\right]
\left(\begin{array}{c}
A_\bot\\ \tilde h_+\\A_\parallel\\ \tilde h_\times
\end{array}\right)=0,
\end{eqnarray}
where now $\Delta_M=\kappa_{\rm eff}\,B_{\rm T}/2$, $\Delta_{\rm g}^\bot=\omega\,(n_{\rm g}^\bot-1)$, $\Delta_{\rm g}^\parallel=\omega\,(n_{\rm g}^\parallel-1)$, and $\Delta^{\bot,\parallel}_{\gamma}$ are defined as in the previous section. In addition, we have more properly defined the amplitude of the GWs polarizations states as $\tilde h_{+,\times}=\kappa/\kappa_{\rm eff}\,h_{+,\times}$, which has allowed us to obtain a symmetric mixing matrix.
Note that system (\ref{Matrix}) describes graviton-photon mixing for ATGs that may predict generically modified propagation and coupling of GWs, but keeping the electromagnetic field minimally coupled to gravity.

\section{Conversion probability}\label{sec:P}
In view of the system (\ref{Matrix}) one can note that there is no mixing between both polarizations. So, as it has been done in reference~\cite{Dolgov:2012be} for the general relativistic case, we can split the system into two similar subsystems defining
\begin{eqnarray}\label{Matrix2}
M_\lambda=
\left({\begin{array}{cc}
   \Delta^\lambda_ {\gamma} & \Delta_M  \\
   \Delta_M & \Delta_{\rm g}^\lambda \\
  \end{array} } \right),\quad
\Psi_\lambda=\left(\begin{array}{c}
A_\lambda\\ \tilde h_\lambda
\end{array}\right),
\end{eqnarray}
where the subindex $\lambda$ denotes the two possible polatizations; however, in what follows, we do not include the sub-index $\lambda$ for simplicity.
As $M$ is a symmetric matrix, it can be diagonalized by a rotation.
Therefore, we have
\begin{equation}
\left[(\omega-i\partial_z)+M\right]\Psi=0=\left[(\omega-i\partial_z)+M'\right]\Psi',
\end{equation}
with $M'=U^TM U$, $\Psi'=U^T\Psi$.
The eigenvalues of $M$ and the angle of rotation are given by 
\begin{eqnarray}
\quad m_{1,2}&=&\frac{1}{2}\left[\Delta_{\rm g}+\Delta_{\gamma}\mp\sqrt{4\Delta_{\rm M}^2+\left(\Delta_{\rm g}-\Delta_{\gamma}\right)^2}\right],\\
\tan(2\theta)&=&\frac{2\Delta_{\rm M}}{\Delta_{\rm g}-\Delta_{\gamma}}.
\end{eqnarray}
So, we obtain a solution for the system with a similar expression to that presented in reference~\cite{Dolgov:2012be}, taking into account our conventions and the different definition of the quantities due to the diagravitational medium. This is
\begin{eqnarray}
\hspace{-2.1cm}A(z)=\left[\cos^2\theta \,e^{-im_1z}+\sin^2\theta \,e^{-im_2z}\right] A(0)
-\sin\theta\cos\theta\left[e^{-im_1z}-e^{-im_2z}\right]\tilde h(0),\\
\hspace{-2cm}\tilde h(z)=-\sin\theta\cos\theta\left[e^{-im_1z}-e^{-im_2z}\right]A(0)
+\left[\sin^2\theta \,e^{-im_1z}+\cos^2\theta \,e^{-im_2z}\right]\tilde h(0),
\end{eqnarray}
where we have absorbed a global phase $e^{i\omega z}$ in $A(z)$ and $\tilde h(z)$. Now, if we assume that initially there are no photons, the probability of graviton-photon conversion is
\begin{equation}\label{P}
P_{{\rm g}\rightarrow{\gamma}}=\frac{\Delta_{\rm M}^2}{\Delta_{\rm M}^2+\left(\Delta_{\rm g}-\Delta_{\gamma}\right)^2/4}
\sin^2\left(\sqrt{\Delta_{\rm M}^2+\left(\Delta_{\rm g}-\Delta_{\gamma}\right)^2/4}\,\cdot z\right).
\end{equation}
The results of the general relativistic mixing are recovered in equation (\ref{P}) for $\Delta_{\rm g}=0$, which degenerate to graviton-photon conversion or resonance if $\Delta_{\gamma}=0$.

\paragraph{Maximum mixing or resonance.} In GR graviton-photon mixing is maximized for $n_{\gamma}=1$. This process corresponds to a maximum mixing angle $\theta=\pi/4$, leading to a probability
\begin{equation}\label{PGR}
P_{{\rm g}\rightarrow{\gamma}}=\sin^2\left(\Delta_{\rm M}\,z\right)=\sin^2\left(\kappa\,B_{\rm T}\,z/2\right).
\end{equation}
Therefore, it is possible to have complete conversion of GWs into EMWs when these propagate through the magnetic field.
In ATGs there is resonance for $\Delta_{\rm g}=\Delta_{\gamma}$, that is $n_{\rm g}=n_{\gamma}$. Then one obtains
\begin{equation}
P_{{\rm g}\rightarrow{\gamma}}=\sin^2\left(\Delta_{\rm M}z\right)=\sin^2\left(\kappa_{\rm eff}\,B_{\rm T}\,z/2\right),
\end{equation}
where we can see that only ATGs inducing an effective gravitational coupling modify the conversion probability  in the very particular limit $\Delta_{\rm g}=\Delta_{\gamma}$.

\paragraph{Weak mixing.} We consider now the case of weak mixing, that is $\theta\ll1$, which corresponds to $(\Delta_{\rm g}-\Delta_{\gamma})^2\gg\Delta_{\rm M}^2$. Then, we have
\begin{eqnarray}\label{Pwm}
P_{{\rm g}\rightarrow{\gamma}}&=&\frac{4\Delta_{\rm M}^2}{\left(\Delta_{\rm g}-\Delta_{\gamma}\right)^2}\sin^2\left(|\Delta_{\rm g}-\Delta_{\gamma}|\,z/2\right).
\end{eqnarray}
This probability reduces to 
\begin{eqnarray}
P_{{\rm g}\rightarrow{\gamma}}&=&\Delta_{\rm M}^2z^2=\kappa_{\rm eff}^2 \,B_{\rm T}^2\,z^2/4,
\end{eqnarray}
for a path even shorter than the oscillation length $l_{\rm osc}=2\pi/|\Delta_{\rm g}-\Delta_{\gamma}|$. Hence, again, only 
ATGs inducing an effective gravitational coupling modify the mixing probability in this regime, amplifying it for $\kappa_{\rm eff}>\kappa$.
It should be noted, however, that any term appearing in the effective gravitational refractive index would be relevant when the propagation path is larger or comparable to the oscillation length, leading to the probability (\ref{Pwm}). 

\section{Discusion}\label{sec:d}
We have investigated the phenomenon of graviton-photon oscillation in ATGs. The alternative gravitational framework that we have considered is quite general. We have taken into account that ATGs may predict that GWs propagate in a different fashion than in GR and that the effective gravitation coupling of these tensor modes differs from $\kappa$. Our study, therefore, applies at the very least to Horndeski theory, ghost free massive gravity and bigravity, and some Lorentz breaking theories. Nevertheless, we have made some important assumptions.
On one hand, we have considered that the electromagnetic field is minimally coupled to gravity. We could expect to have a different dynamical graviton-photon system, for example, in theories with a non-minimal curvature-matter coupling~\cite{Bertolami:2017svl}. An interesting issue would be to investigate how the mixing probability translates when the predictions of the theory under study are equivalent to those of a minimally coupled frame.
On the other hand, we have assumed that the connection is metric-compatible. One should explore whether similar results could be obtained in Palatini theories~\cite{Jimenez:2015caa}.

In this framework, we have obtained the probability of graviton to photon conversion in the presence of a magnetic field, as this situation can describe graviton-photon mixing due to primordial GWs crossing cosmic magnetic fields. This phenomenon can produce, therefore, observable effects in the electromagnetic cosmic background.   As the conversion probability depends on the effective gravitational coupling and on the effective gravitational refractive index, modifications of the general relativistic predictions could be expected if an ATG were the theory describing our Universe. In the particular regimes that we have explicitly considered, the conversion would be amplified if $\kappa_{\rm eff}>\kappa$. It is even more relevant to note that the effective gravitational refractive index is a function of the frequency, having a different functional form for different ATGs. Therefore, whereas for GR the most efficient sources of GWs for graviton-photon mixing are primordial black holes and, therefore, the produced EMWs could be relevant in the cosmic X-ray background  \cite{Dolgov:2012be}, the conversion probability will be maximized for different GW-sources depending on the gravitational theory. So, the particular frequency band of the cosmic electromagnetic background that could have an extra contribution due to this mixing should be calculated for each theory.  

Furthermore, it is worth noting that modifications of the general relativistic predictions can be significant if the mixing took place during the early stages of the cosmic evolution, when primordial magnetic fields should be stronger than now and the constraints on the gravitational coupling and the speed of GWs are less restrictive than those for the current epoch. 
On the other hand, as we have mentioned in the introduction, the recent direct measurements of GWs have put strong constraints on the posible modification of the propagation of GWs in the recent Universe~\cite{Lombriser:2015sxa,Lombriser:2016yzn,Ezquiaga:2017ekz,Creminelli:2017sry,Sakstein:2017xjx,Baker:2017hug}, imposing in particular $c_T(t_{\rm now})\simeq1$. However, modifications of the general relativistic predictions for the conversion process may also be relevant if the mixing took place recently for ATGs having any nontrivial term in the effective gravitational refractive index $n_{\rm g}$ and/or $\kappa_{\rm eff}\neq\kappa$ (but with with $c_T(t_{\rm now})\simeq1$). Whereas at recent times one should expect to have weaker cosmic magnetic fields, magnetic fields of astrophysical origin may play a relevant role in this case. So, new constraints on ATGs or signatures of modifications of GR may be obtained when contrasting the predictions of a particular ATG with the observations of the electromagnetic cosmic background.

It should be emphasized that our results are the first attempt to describe the process of graviton-photon oscillation in ATGs. Some possible extensions of our treatment could take into account the backreaction of the additional degrees of freedom used to formulate the ATG and the potential coupling of their perturbations to the tensor perturbations. In addition, we have assumed that the effective gravitational refractive index and coupling can be treated as constant parameters along the intervals of interest. Nevertheless, some non-trivial effect may emerge in regions of large variation of the effective refractive index.  Finally, it should be noted that our formalism may also be extended to take into account potential effects due to lost of coherence.

\section*{Acknowledgments}
We thank Douglas Singleton for useful comments.
This work is supported by the project FIS2016-78859-P (AEI/FEDER, UE). PMM gratefully acknowledges financial support provided through the Research Award L'Or\'eal-UNESCO FWIS (XII Spanish edition).


\end{document}